# Preseismic electric field "strange attractor" like precursor analysis applied on large (Ms > 5.5R) EQs, which occurred in Greece during December 1st, 2007 - April 30th, 2008.

Thanassoulas[1], C., Klentos[2], V., Verveniotis, G.[3], Zymaris, N.[4]

1. Retired from the Institute for Geology and Mineral Exploration (IGME), Geophysical Department, Athens, Greece.
   e-mail: thandin@otenet.gr - URL: www.earthquakeprediction.gr

2. Athens Water Supply & Sewerage Company (EYDAP),
   e-mail: klenvas@mycosmos.gr - URL: www.earthquakeprediction.gr

3. Ass. Director, Physics Teacher at 2nd Senior High School of Pyrgos, Greece.

4. Retired, Electronic Engineer.
   e-mail: nik.zym@tellas.gr

**Abstract.**

In order to investigate the capability of the preseismic electric field "strange attractor like" precursor as a time predictor of a large EQ within a short time window (short-term prediction), the specific methodology was applied on the Earth's electric field recorded during a rather long seismically active period (December 1st, 2007 - April 30th, 2008) of Greece. During this period of time a number (**8**) of large (**Ms > 5.5R**) earthquakes took place. The particular analysis is presented in detail for the following EQs: the Monemvasia EQ (January 6th 2008, **Ms = 6.6R**), the Methoni EQs (February 14th 2008 **Ms = 6.7R**, February 19th 2008 **Ms = 5.6R**, February 20th 2008 **Ms = 6.5R**, February 26th 2008 **Ms = 5.7R**), the Skyros EQ (March 19th 2008 **Ms = 5.5R**) and the Mid Southern Creta EQ (March 28th 2008 **Ms = 5.6R**). The obtained results from the analysis of the afore mentioned EQs, in conjunction to the ones obtained from an earlier presentation of the particular methodology (Thanassoulas et al. 2008a), suggest: an average time of initiation of the preseismic precursor of the order of (**9**) days before the EQ occurrence, a precursor average duration of the order of (**7**) days, while the elapsed time between the end of the precursor till the EQ occurrence time is, at average, only (**2**) days. These results suggest an objective and easy to apply method for the short-term EQ prediction, based on the registration of the Earth's electric field on ground surface.

## 1. Introduction.

The "mapping technique" which is used in the study of the "non-linear" dynamic systems (Nusse and Yorke 1998, Korsch et al. 2008) was applied on the Earth's electric field recorded at two distant monitoring sites (Thanassoulas et al. 2008a). The results of the analysis of the Earth's electric field, for some days before the occurrence of three large earthquakes, using this technique, indicated that the corresponding daily compiled phase maps exhibit a peculiar behaviour. Very briefly, long before the EQ occurrence the phase map is characterized by random in direction hyperbolas, as long as the recorded Earth's electric field, registered at the two monitoring sites, is not correlated each other, followed by a period of some days in which ellipses characterize the phase map, suggesting a degree of interrelationship between the Earth's electric fields recorded at both monitoring sites, and finally follows a short period, of some days, where the phase map is again characterized by hyperbolas and then follows the EQ occurrence.

Three such examples of the specific analysis have been presented to date (Thanassoulas 2007, Thanassoulas et al. 2008a). These examples were chosen in random, due to their large EQ magnitude, from the list of the earthquakes that took place recently in Greece. The time elapsed between each other is of the order of a few months. Therefore, the question that immediately rises is: how this methodology behaves in a continuous longer (of some months) period of time when consecutive large EQs take place? In order to answer this question, a time period of five months (from December 1st, 2007 to April 30th, 2008) was considered in which a quite large number (**8**) of EQs (**Ms > 5.5R**) took place in Greece. The phase map was compiled on a daily basis from the recordings of the Earth's electric field registered by **PYR** and **HIO** monitoring sites. The detailed analysis is as follows.

## 2. The data.

The time period which will be analyzed spans from December 1st, 2007 to April 30th, 2008. During this period of time (five months) an unusual large number (**8**) of strong (**Ms > 5.5R**) EQs took place that produced a social anxiety as well as in seismological community too. These EQs are presented in the following map depicted in figure (**1**). The background of the map is the seismic potential calculated for the Greek territory for the year 2000 (Thanassoulas et al. 2008) while the specific EQs are presented by blue circles. Numbering of the EQs links them with the following **Table – 1** where the date of occurrence and the corresponding magnitude have been tabulated. The monitoring sites of the Earth's electric field (**PYR, ATH** and **HIO**) are denoted with red lettering.

**TABLE – 1**

| 1 | Monemvasia EQ, January 6th, 2008, Ms = 6.6R | 2 | Methoni EQ, February 20th, 2008, Ms = 6.5R |
|---|---|---|---|
| 5 | SW of Patras EQ, February 2nd, 2008, Ms = 5.5R | 2 | Methoni EQ, February 26th, 2008, Ms = 5.7R |
| 2 | Methoni EQ, February 14th, 2008, Ms = 6.7R | 3 | Skyros EQ, March 19th, 2008, Ms = 5.5R |
| 2 | Methoni EQ, February 19th, 2008, Ms = 5.6R | 4 | Mid. S. Creta EQ, March, 28th, 2008, Ms = 5.6R |

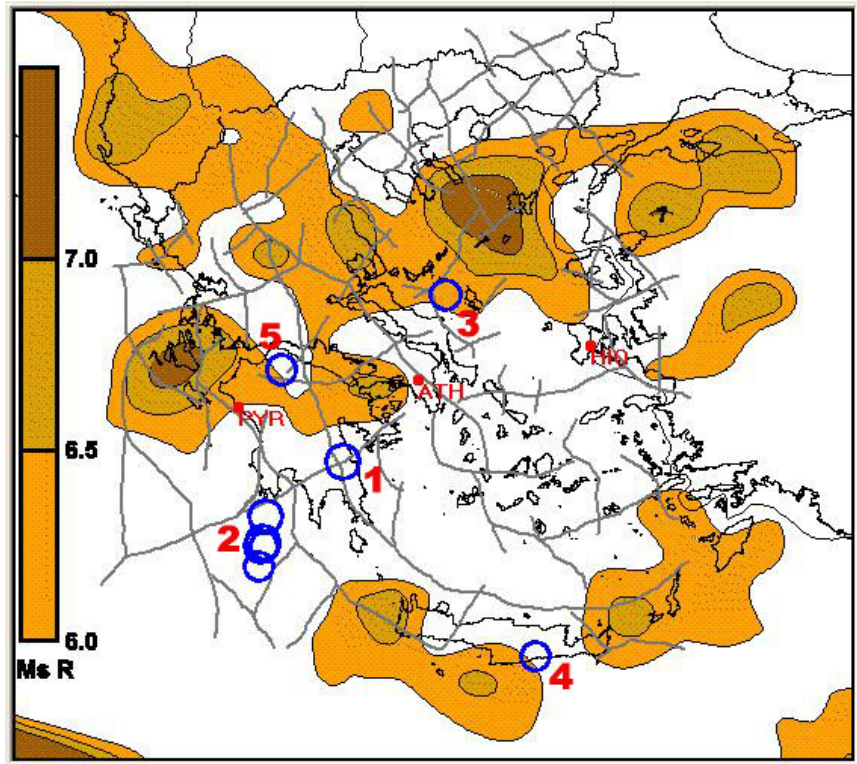

Fig. 1. Large (blue circles, **Ms > 5.5R**) EQs that occurred in Greece during the time period from December 1st, 2007 to April 30th, 2008. For details see **Table – 1.**

During the same period of time, the electric field of the Earth, which was recorded by **PYR** and **HIO** monitoring sites, after normalization (to a unit length of orthogonal **N - S, E – W** pair of dipoles) and extracting its monochromatic component of **T = 24h**, is presented in the following figure (**2**) for the **PYR** and figure (**3**) for the **HIO** monitoring sites.

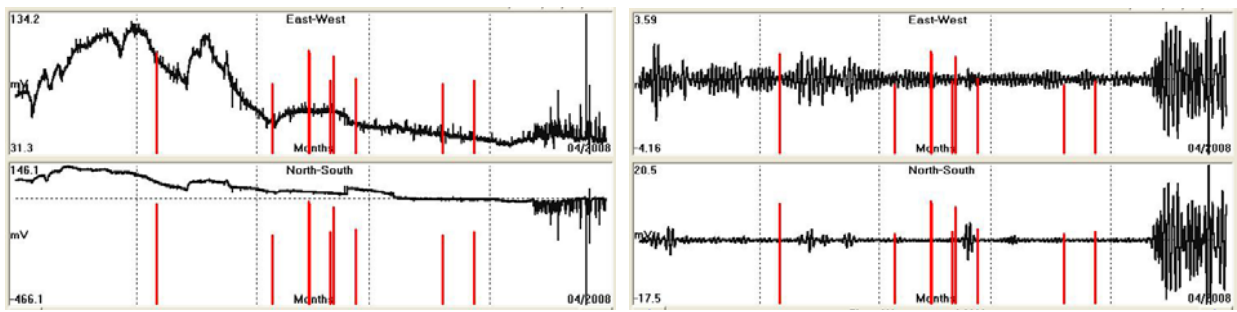

Fig. 2. The Earth's electric field recorded for the time period from December 1st, 2007 to April 30th, 2008 by **PYR** monitoring site. Left = normalized raw data. Right = extracted monochromatic (**T = 24h**) data. The red bars indicate the time of occurrence and magnitude of the EQs (**Ms > 5.5R**) which took place in the same period of time.

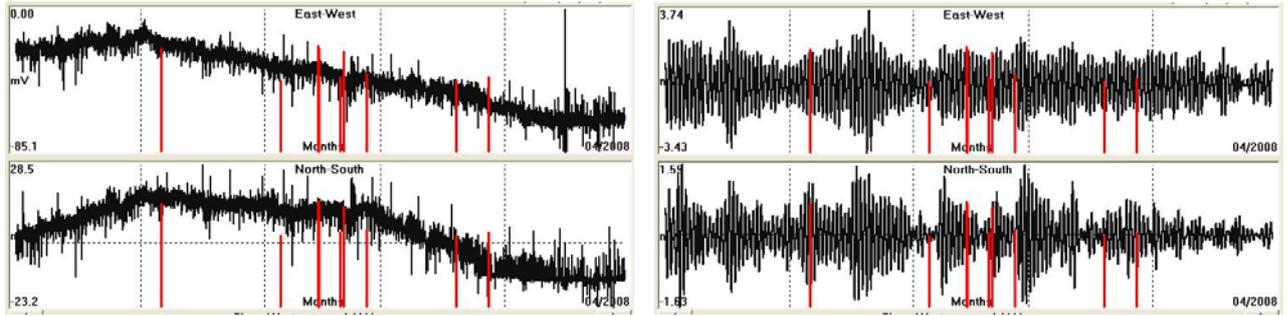

Fig. 3. The Earth's electric field recorded for the time period from December 1$^{st}$, 2007 to April 30$^{th}$, 2008 by **HIO** monitoring site. Left = normalized raw data. Right = extracted monochromatic (**T = 24h**) data. The red bars indicate the time of occurrence and magnitude of the EQs (**Ms > 5.5R**) which took place in the same period of time.

### 3. Examples presentation.

Phase maps were compiled, for all these EQs on a daily basis, for a period of time: quite before, close to and after the EQ occurrence. The results are collectively presented in bar graphs where the observed ellipses (blue bars) per day are presented vs. time (in days) and the day of occurrence of the corresponding EQ (red bar). A zero value on a specific day means that no ellipses were generated but only hyperbolas were identified. Moreover, typical examples of the generated corresponding phase map are presented for the time periods: quite before, close to and after the EQ occurrence in order to show the peculiar behavior of the "strange attractor like" seismic precursor. The corresponding EQ, for each analyzed period of time, is denoted in the phase maps by blue concentric circles.

### 3.1. Monemvasia EQ (January 6$^{th}$ 2008, Ms = 6.6R).

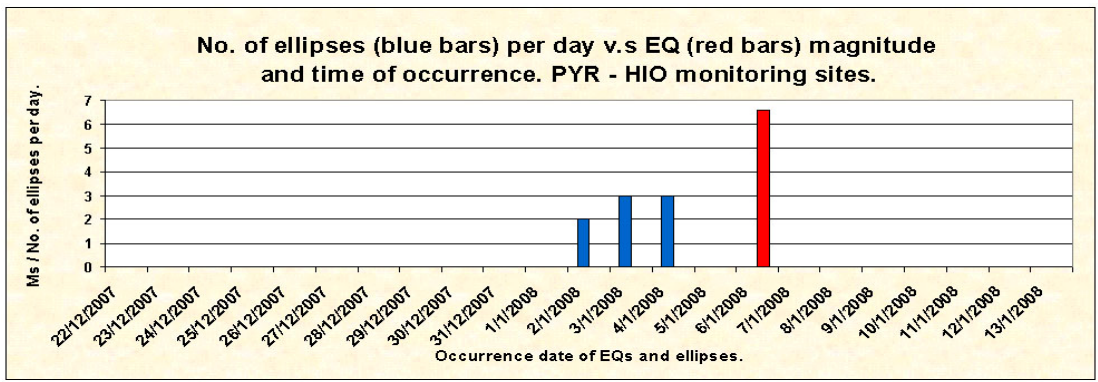

Fig. 4. Preseismic "strange attractor like" precursor (ellipses - blue bars) observed prior to the Monemvasia EQ (red bar, January 6$^{th}$ 2008, Ms = 6.6R).

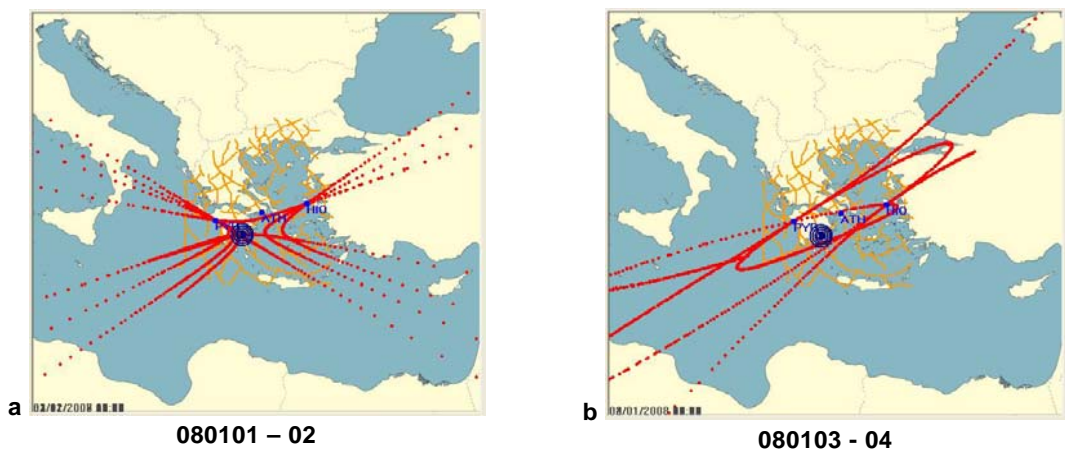

| a | b |
| --- | --- |
| 080101 – 02 | 080103 - 04 |

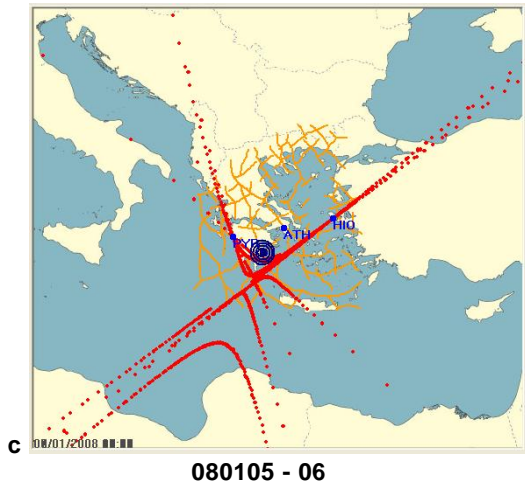

**080105 - 06**

Fig. 5. Examples (**a, b, c**) of phase maps generated for the corresponding dates (yymmdd) for the Monemvasia EQ (blue circles, January 6th 2008, Ms = 6.6R).

**3.2**. **Methoni EQs (February 14th 2008 Ms = 6.7R, February 19th 2008 Ms = 5.6R, February 20th 2008 Ms = 6.5R, February 26th 2008 Ms = 5.7R).**

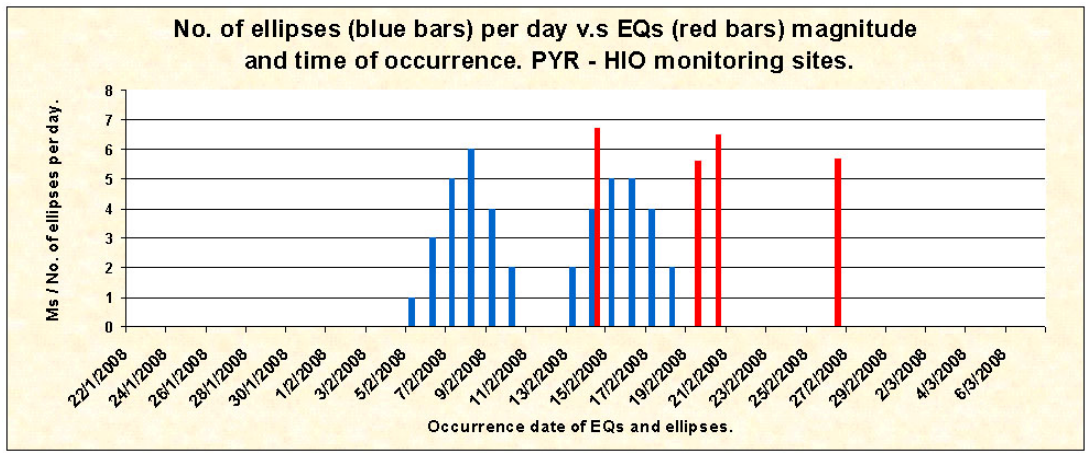

Fig. 6. Preseismic "strange attractor like" precursor (ellipses - blue bars) observed prior to the Methoni EQs (red bars, February 14th 2008 Ms = 6.7R, February 19th 2008 Ms = 5.6R, February 20th 2008 Ms = 6.5R, February 26th 2008 Ms = 5.7R).

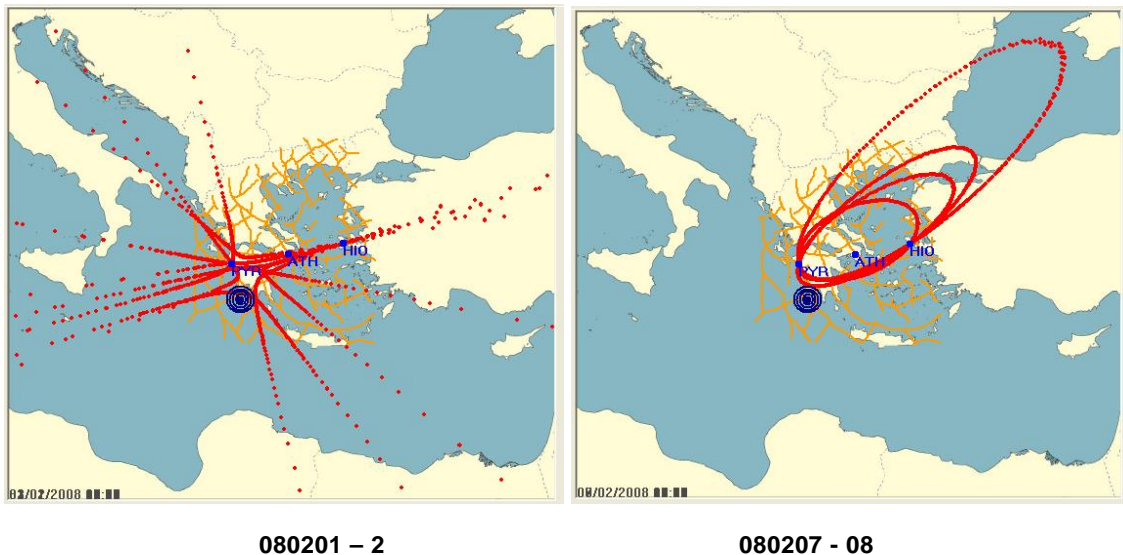

**080201 – 2**          **080207 - 08**

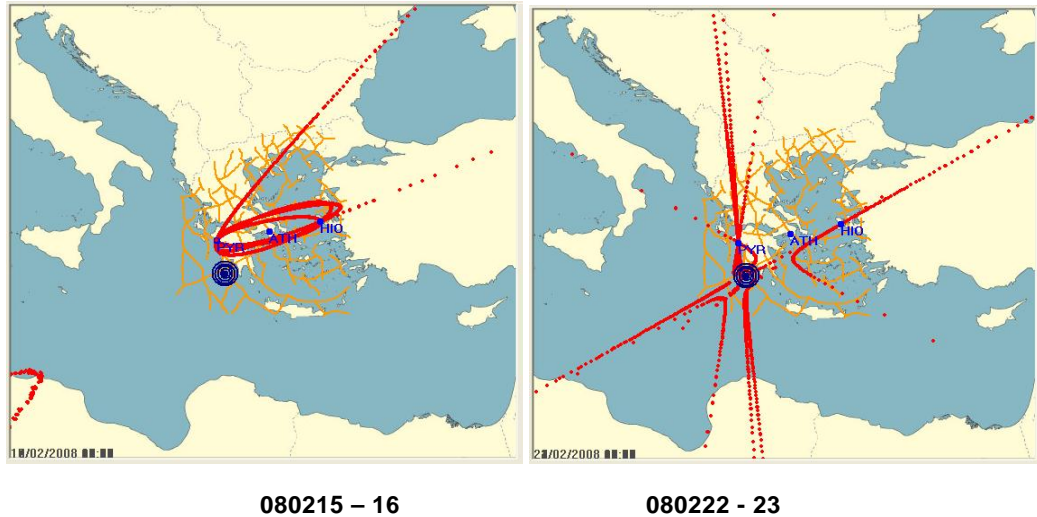

| 080215 – 16 | 080222 - 23 |

Fig. 7. Examples of phase maps generated for the corresponding dates (yymmdd) for the Methoni EQ (blue circles, February 14[th] 2008, Ms = 6.7R).

### 3.3. Skyros EQ (March 19[th] 2008 Ms = 5.5R).

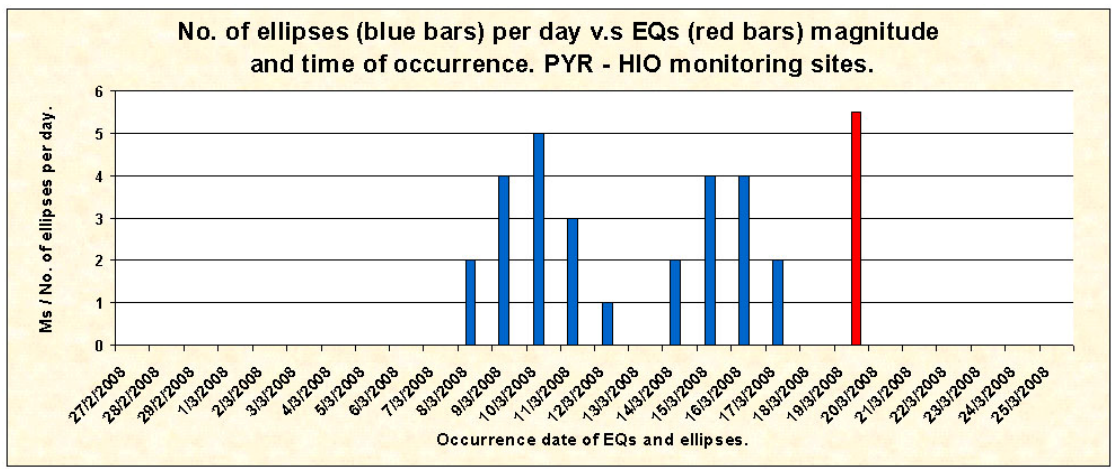

Fig. 8. Preseismic "strange attractor like" precursor (ellipses - blue bars) observed prior to the Skyros EQ (red bar, March 19[th] 2008, Ms = 5.5R).

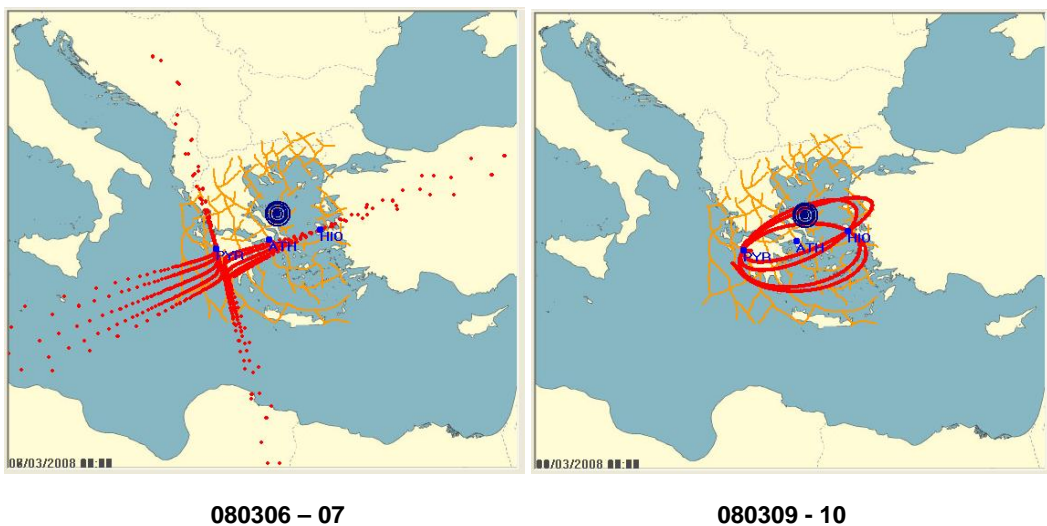

| 080306 – 07 | 080309 - 10 |

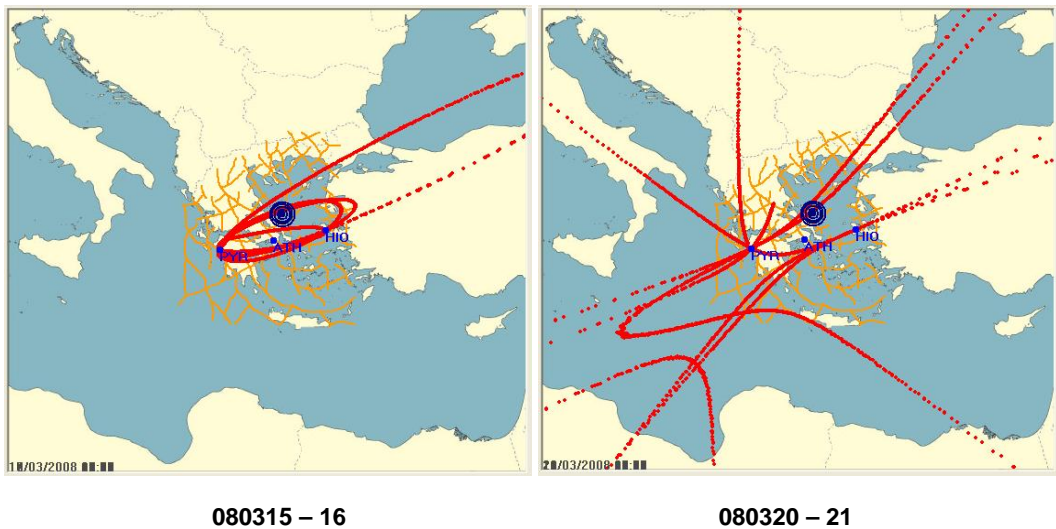

**080315 – 16**   **080320 – 21**

Fig. 9. Examples of phase maps generated for the corresponding dates (yymmdd) for the Skyros EQ (blue circles, March 19[th] 2008 Ms = 5.5R).

### 3.4. Mid. Southern Creta EQ (March 28[th] 2008 Ms = 5.6R).

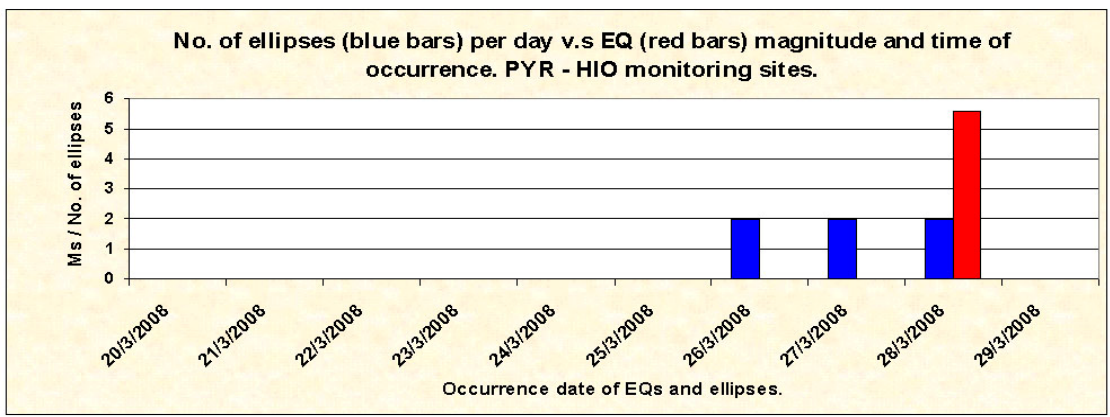

Fig. 10. Preseismic "strange attractor like" precursor (ellipses - blue bars) observed prior to the Mid. Southern Creta EQ (red bar, March 28[th] 2008 Ms = 5.6R).

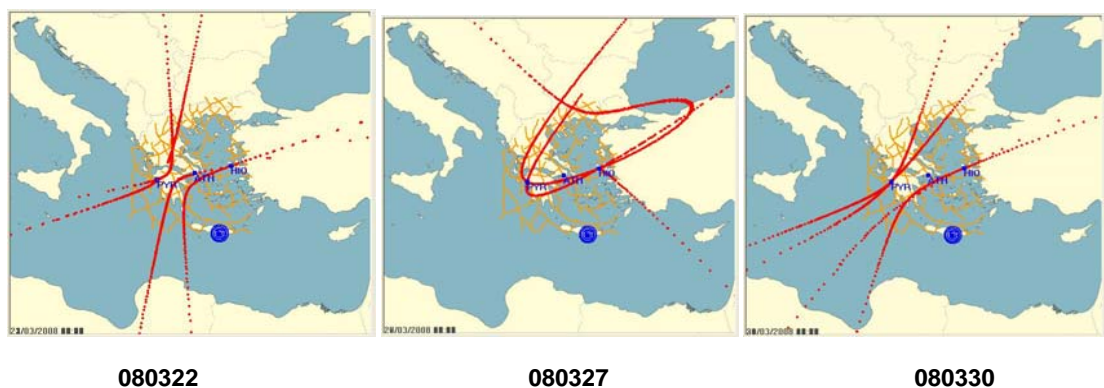

**080322**   **080327**   **080330**

Fig. 11. Examples of phase maps generated for the corresponding dates (yymmdd) for the Mid. Southern Creta EQ (blue circles, March 28[th] 2008 Ms = 5.6R).

## 4. Discussions – conclusions.

The analysis of the obtained results shows some interesting features that concern the "short-term", in terms of time-window, earthquake prediction. Six (**6**) EQs out of eight (**8**) were preceded by the "strange attractor like" precursor. This corresponds to a **75%** hit success which is quite a large value. However, one EQ (SW of Patras, no. 5) was not preceded by any "strange attractor like" precursor at all, that is the EQ of February 4$^{th}$ 2008, Ms = 5.5R, while the second failure refers to the EQ (Methoni seismogenic area, no.2) of February 26$^{th}$ 2008, Ms = 5.7R. Probably, the later, as being a large aftershock of the main seismic events of Ms = 6.7R and 6.5R (no. 2) did not produce any electric signal due to the fact that the most of the stored in the seismogenic area strain energy had been released. The EQs of: Monemvasia, (no.1), January 6$^{th}$, 2008, Ms = 6.6R, Skyros (no.3), March 19$^{th}$, 2008, Ms = 5.5R and Mid. S. Creta (no.4), March, 28$^{th}$, 2008, Ms = 5.6R, due to the fact that they were isolated, in time, seismic events, generated clear and observable "strange attractor like" preseismic precursors.

Properties of the "strange attractor like" seismic precursor had been presented by Thanassoulas et al. (2008a). These properties referred to **a)** the time (in days) of initiation of the precursor before the occurrence of the EQ, **b)** the duration (in days) of the precursor and **c)** the elapsed time (in days) between the end of the precursor and the time of EQ occurrence. The same procedure will be followed for the current data set and only for the EQs which were preceded by a "strange attractor like" precursor. These properties are tabulated in **Table – 2**. In this table are tabulated: **(a)** the elapsed time (in days) from the initiation of the precursor till the EQ occurrence, **(b)** the duration of the precursor in days, and **(c)** the time elapsed (in days) after the vanishing of the precursor till the time of occurrence of the EQ.

**TABLE – 2**

| Earthquake (date, magnitude) | Precursor initiation (days) | Precursor duration (days) | Precursor end – EQ occur. Time (days) |
|---|---|---|---|
| January 6$^{th}$ 2008, Ms = 6.6R | 9 | 6 | 3 |
| February 14$^{th}$ 2008, Ms = 6.7R | 10 | 6 | 4 |
| February 19$^{th}$ 2008, Ms = 5.6R | 7 | 6 | 1 |
| February 20$^{th}$ 2008, Ms = 6.5R | 8 | 6 | 2 |
| March 19$^{th}$ 2008, Ms = 5.5R | 12 | 10 | 2 |
| March 28$^{th}$ 2008, Ms = 5.6R | 3 | 3 | 0 |
| Average values (integer days): | 8 | 6 | 2 |

At this point it is worth to recall the first results (**Table – 3**) of a similar work presented by Thanassoulas et al. (2008a).

**TABLE – 3**

| Earthquake (date, magnitude) | Precursor initiation (days) | Precursor duration (days) | Precursor end – EQ occur. Time (days) |
|---|---|---|---|
| April 19$^{th}$, 2007, Ms = 5.4R | 15 | 13 | 2 |
| June 8$^{th}$, 2008, Ms = 7.0R | 11 | 9 | 2 |
| June 21$^{st}$, 2008, Ms = 6.0R | 8 | 5 | 3 |
| Average values (integer days): | 11 | 9 | 2 |

A comparison of **Table – 3** to **Table – 2** shows that the average values of the precursor signal initiation and duration, shown in **Table – 3**, were overestimated (probably due to the small statistical dataset) at almost 30% from the corresponding values calculated from **Table – 2**. However, the average elapsed time from the end of the preseismic precursor to the time of EQ occurrence is the same in both tables. The later suggests that the remaining time from the end of the precursor to the occurrence of the EQ is more stable than the time of initiation or duration of the very same precursor in the frame of this analyzed data set.

Since large EQs are not an event that takes place quite frequently, it poses a difficulty in earthquake prediction studies. Although Greece is a highly seismogenic region, at average, a large EQ of **Ms > 6.0R** takes place every 1 – 1.5 years. Therefore, studies on precursors of the present type, evidently, require a dataset that extends largely in the past. A small substitute and quick approach to this is to combine the two different data sets in one. This is valid since the two data sets consist of different large EQs which took place in the Greek territory during the time period of 2007 – 2008. The combined data set is presented in the following **Table – 4.**

**TABLE – 4.**

| Earthquake (date, magnitude) | Precursor initiation (days) | Precursor duration (days) | Precursor end – EQ occur. Time (days) |
|---|---|---|---|
| April 19$^{th}$, 2007, Ms = 5.4R | 15 | 13 | 2 |
| January 6$^{th}$, 2008, Ms = 6.6R | 9 | 6 | 3 |
| February 14$^{th}$, 2008, Ms = 6.7R | 10 | 6 | 4 |
| February 19$^{th}$, 2008, Ms = 5.6R | 7 | 6 | 1 |
| February 20$^{th}$, 2008, Ms = 6.5R | 8 | 6 | 2 |
| March 19$^{th}$, 2008, Ms = 5.5R | 12 | 10 | 2 |
| March 28$^{th}$, 2008, Ms = 5.6R | 3 | 3 | 0 |
| June 8$^{th}$, 2008, Ms = 7.0R | 11 | 9 | 2 |
| June 21$^{st}$, 2008, Ms = 6.0R | 8 | 5 | 3 |
| Average values (integer days): | 9 | 7 | 2 |

The average values of **Table – 4** resemble quite well those of **Table – 2**.

The fact that the Greek territory exceeded its average value of seismic activity expected within a year (1-1.5 EQs of **M > 6R** per year) during the time period of this experiment, helped us a lot to obtain valuable data for testing this methodology. Within the total time period of study, five (**5**) large EQs occurred of **Ms > 6.0R**. Hence, it is worth, and interesting too, to apply this methodology on a data set with a minimum threshold of EQ magnitude of **Ms >= 6.0R**. This is demonstrated in the following **Table – 5.**

TABLE – 5.

| Earthquake (date, magnitude) | Precursor initiation (days) | Precursor duration (days) | Precursor end – EQ occur. Time (days) |
|---|---|---|---|
| January 6$^{th}$, 2008, Ms = 6.6R | 9 | 6 | 3 |
| February 14$^{th}$, 2008, Ms = 6.7R | 10 | 6 | 4 |
| February 20$^{th}$, 2008, Ms = 6.5R | 8 | 6 | 2 |
| June 8$^{th}$, 2008, Ms = 7.0R | 11 | 9 | 2 |
| June 21$^{st}$, 2008, Ms = 6.0R | 8 | 5 | 3 |
| Average values (integer days): | 9 | 6 | 3 |

Practically, and in terms of short-term earthquake prediction, the average values are very similar to those shown in **Table – 4**. Apart from the increase at a day's time of the "end of precursor – time of EQ occurrence" elapsed time period, generally the sequence of the precursory phenomenon is as follows: precursor initiation starts some days before the EQ, followed by a precursor duration of a slightly shorter period of time and then follows a very short period of 2 – 3 days, before the EQ occurrence, when the precursor vanishes.

Furthermore, if the entire present data set is combined to the one presented by Thanassoulas et al. (2008a), then the total number of the EQs that took place in the whole study period of time (April 2007 – June 2008) becomes (**11**) while only two (**2**) of them did fail to generate a "strange attractor like" precursor. This corresponds to an **81.8%** precursor success compared to **75.0%** success of the present work data set.

In conclusion, the presented examples show that the "strange attractor like" seismic precursor was capable to provide satisfactory results, so far, concerning the time of occurrence of large EQs (**Ms > 5.5R**) for the study period in the Greek seismogenic area. It is clear that this method should be tested further more for longer time periods (of some years) and for a larger number of large EQs, in order to represent a valid statistical test, provided that Earth's electric field data have been registered for a similar long period too.

Finally, for those who are interested in the practical – technical details of it, the distance between **PYR** and **HIO** monitoring sites is 423 Km, while the lengths of the receiving electric dipoles range from 160m (for **PYR** monitoring site) to 200m (for **HIO** monitoring site). A full and detailed technical description of the used hardware and electrical signals analysis methodology was presented by Thanassoulas (2007).

## 5. References.